\documentclass[a4paper,11pt]{article}
\usepackage{graphicx,amssymb,amstext,amsmath,mathabx,mathtools,esvect}

\usepackage{floatflt}
\usepackage{subcaption}
\usepackage[dvipsnames]{xcolor}

\definecolor{cbl}{rgb}{0,0,1}                

\newcommand{\bc}{\begin{center}}
\newcommand{\ec}{\end{center}}
\def\ba#1{\begin{array}{#1}\displaystyle}
\newcommand{\ea}{\end{array}}

\newcommand{\beq}{\begin{equation}}
\newcommand{\eeq}{\end{equation}}
\newcommand{\beqa}{\begin{eqnarray}}
\newcommand{\eeqa}{\end{eqnarray}}

\newcommand{\bi}{\begin{itemize}}
\newcommand{\ei}{\end{itemize}}

\newcommand{\bra}{\langle}
\newcommand{\ket}{\rangle}

\newcommand{\Tr}{{\rm Tr}}

\newcommand{\TT}{{\cal T}}

\usepackage{cite}

\definecolor{purple_nice}{rgb}{0.4,0.2,0.7}
\definecolor{fuel_blue}{RGB}{42,162,185}
\definecolor{YInMn_blue}{RGB}{46, 80, 144}
\definecolor{ultramarine}{RGB}{63, 0, 255}
\definecolor{KLEIN_blue}{rgb}{0, 0.18, 0.65}
\usepackage[linktocpage=true]{hyperref}
\hypersetup{
    colorlinks=true,
    linkcolor=YInMn_blue,
    citecolor=ultramarine,
    filecolor=fuel_blue,
    urlcolor=KLEIN_blue,
}

\begin{document}
\begin{titlepage}
\vspace{0.2cm}
\begin{center}

{\large{\bf{Temporal Entanglement in Quantum Field Theory}}}

\vspace{0.8cm} 
{\large\text{Olalla A. Castro-Alvaredo}}

\vspace{0.8cm}
Department of Mathematics, City St George's, University of London,\\ 10 Northampton Square EC1V 0HB, UK\\

\end{center}

\vspace{1cm}

In this paper I propose a branch point twist field approach to computing the {\it temporal entropy}, that is, an entanglement measure across different time regions, as opposed to the usual spacial measures. Considering only the ground state of a gapped theory, I discuss how the shift to time-dependence manifests in form factor calculations. I show that the generalization of the spacial measures to temporal ones reproduces expected features of the temporal entanglement: the entropy is complex, oscillatory and reminiscent of the evolution of entanglement following a global quench. Considering the temporal von Neumann entropy, I argue that spacial and temporal entropies are two sides of the same coin. They both encapsulate universal information about the theory, in particular its mass spectrum. Also in both cases, a quasiparticle picture can be employed to interpret results. This picture connects features of the temporal entropy with those of the spacial entropy following a quantum quench. 

\medskip
\medskip
\medskip

\noindent {\bfseries Keywords:}  Integrability,  Branch Point Twist Fields, Temporal Entanglement
\vfill

\noindent 
o.castro-alvaredo@city.ac.uk\\

\hfill \today

\end{titlepage}
\section{Introduction}
 The aim of this paper is to present a quantum field theoretical study of a measure of {\it temporal entanglement} in 1+1 dimensions. This measure, known as {\it temporal entropy}, was first introduced in \cite{Carignano:2023xbz,Bou-Comas:2024pxf} and it has some unusual entropic features, such as taking complex values. The  temporal entropy is a very natural generalization of the standard spacial measure with the roles of space and time interchanged. In the quantum field theory picture that I develop in this paper this corresponds to replacing spacelike by timelike correlators or to considering correlations across time (dynamical correlators) rather than across space.  The fundamental idea advocated in the original and subsequent papers \cite{Carignano:2023xbz,Bou-Comas:2024pxf,Carignano:2024jxb,Carignano:2025egi} is that while standard {\it spacial measures answer the question of how entangled regions in space are, temporal measures quantify entanglement ``across different times"}. 
\medskip

I will start by setting up the approach that is traditionally employed when studying entanglement measures in 1+1D integrable quantum field theory (IQFT). It is an approach based on relating (spacial) entanglement measures to correlation functions of symmetry fields known as branch point twist fields (BPTFs) \cite{entropy}.
Noether's theorem states that if a system has a continuous symmetry, then there is a corresponding conserved quantity \cite{Noether_1971}. In quantum field theory, the theorem can be translated into a statement about fields: a symmetry gives rise to a conserved current and a conserved charge and this charge generates a symmetry transformation of quantum fields. BPTFs are special quantum fields that implement a symmetry locally in spacetime. In 1+1 dimensions, they sit at the origin of branch cuts and a local symmetry transformation is generated when local fields ``cross" the branch cut. It is in this sense that we say BPTFs are symmetry fields. They play an important role in many contexts, from the computation of partition functions in non-trivial Riemann surfaces \cite{DiMS}, to the field content of orbifolded theories \cite{kniz,orbifold,Borisov,Bouwknegt}  and the definition of order and disorder parameters/fields in statistical models/quantum field theories \cite{Zuber,Schroer,YZam}. See \cite{RevTwist} for a recent review.  This paper exploits the application of BPTFs to the computation of entanglement measures. 

\medskip

The relationship between entanglement measures, partition functions in multi-sheeted Riemann surfaces and BPTFs has been known for some time \cite{CallanW94,HolzheyLW94,Calabrese:2004eu,entropy} and is often referred to as the ``replica trick". The idea is that the QFT partition function on an $n$-sheeted Riemann surface is proportional to the trace of the $n$th power of the reduced density matrix $\rho_A$ associated to a (spacial) subsystem of the theory. We call this subsystem $A$ and its complement $\bar{A}$. If the state of theory is pure, that is, is a vector $|\Psi\ket$, rather than an ensemble, then we have that $\rho_A:=\Tr_{\rm \bar{A}}(|\Psi\ket \bra \Psi|)$. The replica trick is the statement $\mathcal{Z}_n \, \propto \, \Tr_{\rm A} \rho_A^n$, where $\mathcal{Z}_n$ is the replica partition function. In the simplest case, subsystem $A$ is simply connected and in one space dimension has some length $r$. This corresponds to the length of a branch cut in the Riemann surface which has the structure of cyclically connected Riemann sheets. This reduced density matrix is directly related to a measure of entanglement known as the $n$th R\'enyi entropy of the state as:
\beq 
S_n=\frac{1}{1-n} \log(\Tr_{\rm A} \rho_A^n)\,. 
\label{re}
\eeq 
The limit when $n \rightarrow 1$ of this quantity gives the von Neumann entropy $S=-\Tr_{\rm A}(\rho_A \log \rho_A)$, another popular measure of entanglement \cite{bennet}. BPTFs provide an alternative representation of the trace $\Tr_{\rm A} \rho_A^n$, namely as a correlation function. Although the above is valid for any pure state $|\Psi\ket$, in this paper I will consider only the ground state of a gapped or massive theory. 

\medskip

This paper is organized as follows: In Section \ref{BPTFs}, I review some basic properties of BPTFs, in particular how the R\'enyi entropies (\ref{re}) of a connected one-dimensional spacial region may be written in terms of a two-point function of such fields. I then propose a temporal version of the R\'enyi entropies (\ref{re}) which amounts to replacing spacelike by timelike separated fields in the correlator. In Section \ref{FFs}, I review the form factor approach to BPTF correlators and propose how existing universal results for the (spacial) R\'enyi entropy of a connected interval are generalized to its temporal version through simple analytic continuation in the subsystem's size.
In section \ref{von}, I present a universal result for temporal von Neumann entropy of IQFTs with diagonal $S$-matrices and one-particle spectrum. In section \ref{Free}, I present an explicit calculation of the temporal R\'enyi entropies in the free fermion theory and show how the result shares common features with the time-dependent spacial R\'enyi entropies after a global quench.  In section \ref{extra} I present a general discussion of the leading large-time asymptotics of the temporal entropy. I conclude in section \ref{conc}.

\section{Spacelike and Timelike Correlators of BPTFs}
\label{BPTFs}
Consider $n$ identical non-interacting copies of 1+1D QFT. Besides the symmetries that the single copy theory may have, the replica model has permutation symmetry under exchange of any of the copies. By Noether's theorem, in such a theory, there exists a local quantum field associated with each element of the permutation group. Let us denote by $\TT(x,t)$ the symmetry field corresponding to the cyclic permutation $\sigma: j \mapsto  j+1 \mod n$ and by $\tilde{\TT}(x,t)$ its hermitian conjugate, corresponding to the inverse permutation $\sigma^{-1}: j \mapsto  j-1 \mod n$, where $j$ is the copy number. Then, we can write 
\beq 
\frac{\mathcal{Z}_n}{(\mathcal{Z}_1)^n}=\frac{\varepsilon^{4\Delta_n} {}_n\bra \Psi|\mathcal{T}(0,0)\tilde{\mathcal{T}}(r,0)|\Psi\ket_n} {{}_n\bra \Psi|\Psi\ket_n}\, \propto\, \Tr_{\rm A}\rho_A^n\,,
\label{cnk}
\eeq 
that is, the normalized partition function can be written as a normalized two-point function of same-time, spacelike separated BPTFs. Here the normalization is such that all functions become 1 when $n\rightarrow 1$ (in this limit, the BPTF becomes the identity field), $|\Psi\ket_n$ is the replica state ($n$ copies of $|\Psi\ket$), $\varepsilon$ is a short-distance cut-off and $\Delta_n$ is the (conformal) dimension of the BPTFs \cite{kniz,orbifold,Borisov,Bouwknegt,Calabrese:2004eu}
\beq 
\Delta_n=\frac{c}{24}\left(n-\frac{1}{n}\right)\,,
\label{dim}
\eeq 
which depends on $n$ and on a constant $c$, the central charge of the conformal field theory that describes the high energy limit of the model. 
The spacial R\'enyi entropy can then be written as
\beq 
S_n(r)=\frac{1}{1-n}\log \left[\frac{\varepsilon^{4\Delta_n} {}_n\bra \Psi|\mathcal{T}(0,0)\tilde{\mathcal{T}}(r,0)|\Psi\ket_n} {{}_n\bra \Psi|\Psi\ket_n}\right]\,.
\label{rentwist}
\eeq 
Employing this picture and its generalizations, a plethora of models and situations have been studied in the literature. In massive IQFT the entanglement entropies of the ground state have been computed using form factor techniques \cite{entropy,review,next,nexttonext}. Also in the ground state, there have been studies of other entanglement measures such as the logarithmic negativity, first in conformal field theory \cite{negativity1,negativity2,negativity3,negativity4}, using the fields above and their generalizations for other elements of the permutation group, and then also for IQFTs \cite{ourneg,davide}. The BPTF technique has also been used for excited states \cite{excited}, for non-unitary theories \cite{BCDLR,Dupic:2017hpb,Bianchini:2015uea} and for symmetry resolved entanglement measures in the ground state \cite{Horvath:2021rjd,Capizzi:2021kys,Castro-Alvaredo:2023wlr} and in excited states \cite{Capizzi:2022jpx,Capizzi:2022nel,Capizzi:2023ksc}. More general BPTFs can be used to compute the entanglement asymmetry \cite{Capizzi:2023yka}. The formalism has been adapted to out-of-equilibrium situations, as shown for instance in \cite{viti1,viti2,DiSalvo:2024wzu,Lencses:2020jjt,DelVecchio:2023xdl,Kiraly:2025tdr}. In all of these applications, the entanglement measures are associated with spacial bipartitions of the system, that is, the quantum system, which is one-dimensional in space, is divided into two parts. 

\medskip

Consider now a temporal version of (\ref{rentwist}), that is simply
\beq 
S_n(t)=\frac{1}{1-n}\log \left[\frac{\varepsilon^{4\Delta_n} {}_n\bra \Psi|\mathcal{T}(0,0) \tilde{\mathcal{T}}(0,t)|\Psi\ket_n} {{}_n\bra \Psi|\Psi\ket_n}\right]\,.
\label{rentime}
\eeq 
The only difference is that the BPTFs now sit at the same space position but at different moments in time. These are the temporal R\'enyi entropies. {\it The simple exchange of the roles of space and time gives rise to an entanglement measure that looks at correlations across time}  \cite{Carignano:2023xbz}. Technically-speaking timelike correlators are harder to compute that space like correlators. As discussed below, the usual form factor expansion of the correlator (\ref{rentwist}) is rapidly convergent for large distance $r$ (it is essentially an Euclidean correlator), whereas the convergence of the correlator (\ref{rentime}) for large $t$ depends on the properties of the form factors. I will discuss this further in the next sections.

\medskip 

Going back to the construction described in the introduction, the reduced density matrix $\rho_A$ is the result of tracing out the degrees of freedom of a spacial region $\bar{A}$. It measures the correlations that arise when the spacial subsystem $A$ forgets about its spacial environment. In the temporal picture, $\bar{A}$ and $A$ are time intervals and a similar construction now based on a transition rather than a density matrix,  is described in detail in \cite{Carignano:2023xbz}. Once can then think of subsystems $A$ and $\bar{A}$ as representing the past and future of the quantum system. This physical picture is closely aligned with that of global quantum quenches, where a quasiparticle picture explains how excitation pairs produced at time zero become entangled at later times \cite{quench1,quench2,quench3,qualba}. We will see later, especially in section \ref{Free} how my results for the temporal R\'enyi entropies of free fermions  are indeed reminiscent of the evolution of entanglement following a global mass quench in the same theory. It is important to mention that 
BPTFs have been recently employed in this context in \cite{Xu:2024yvf}, where a related quantity, the so-called timelike entropy \cite{Doi:2022iyj,Doi:2023zaf} was computed for conformal field theories. This quantity is related but different from our definition in that it combines timelike and spacelike correlators, and their derivatives. However, no attempt has been made yet to formulate a BPTF construction in massive theories. 
Similarly, the temporal entropy defined in \cite{Carignano:2023xbz,Bou-Comas:2024pxf,Carignano:2024jxb,Carignano:2025egi} is based on so-called transition matrices and it is reminiscent of proposals to compute the entanglement entropy in non-unitary models, where the left and right eigenvectors of the hamiltonian are different from each other, giving rise to a reduced density matrix that is no longer positive-definite \cite{BCDLR,Dupic:2017hpb}. 

\section{Temporal R\'enyi Entropies and Form Factor Computations}
\label{FFs}
In this section I will review the main ideas involved in the form factor construction of correlation functions. Since there is already a vast amount of literature for spacial measures, I will present only a brief summary of the results that I need, see \cite{entropy,nexttonext,review} for more details. Consider a local field $\mathcal{O}(x,t)$ in a general (non-replicated) IQFT. The two-point function in the ground state $|0\ket$ admits the following expansion:
\beq 
\bra 0|\mathcal{O}(0,0)\mathcal{O}(x,t)|0\ket = \sum_{k=0}^\infty \frac{1}{k!}\left[\prod_{i=1}^k \int_{-\infty}^\infty \frac{d\theta_i}{2\pi}\right] |F_k^{\mathcal{O}}(\theta_1,\ldots,\theta_k)|^2 e^{-i t \sum_{j=1}^k e(\theta_j)+i x \sum_{j=1}^k p(\theta_j) }\,,
\label{6}
\eeq 
while its logarithm admits a similar-looking cumulant expansion
\beq 
\log\left[\frac{\bra 0|\mathcal{O}(0,0)\mathcal{O}(x,t)|0\ket}{\bra 0|\mathcal{O}|0\ket^2} \right]= \sum_{k=1}^\infty \frac{1}{k!}\left[\prod_{i=1}^k \int_{-\infty}^\infty \frac{d\theta_i}{2\pi}\right] H_k^{\mathcal{O}}(\theta_1,\ldots,\theta_k) e^{-i t \sum_{j=1}^k e(\theta_j)+i x \sum_{j=1}^k p(\theta_j) }\,.
\label{7}
\eeq 
Here $\theta_i$ are rapidity variables which parametrize the energy $e(\theta)=m\cosh\theta$ and momentum $p(\theta)=m\sinh\theta$ of a particle of mass $m$. The functions $F_k^{\mathcal{O}}(\theta_1,\ldots,\theta_k)$ are the form factors of a $k$-particle state, which can be formally written as:
\beq 
F_k^{\mathcal{O}}(\theta_1,\ldots,\theta_k):=\bra 0|\mathcal{O}(0,0)|\theta_1,\ldots,\theta_k\ket\,.
\eeq 
whereas the cumulants $H_k^{\mathcal{O}}(\theta_1,\ldots,\theta_k)$ are specific combinations of the form factors (see e.g. \cite{karo}) which can be worked out systematically by comparing the expansions (\ref{6}) and (\ref{7}). The first two functions are:
\beq 
(F_0^{\mathcal{O}})^2 H_1^{\mathcal{O}}(\theta)=|F_1^{\mathcal{O}}(\theta)|^2\qquad \mathrm{and} \qquad (F_0^{\mathcal{O}})^4 H_2^{\mathcal{O}}(\theta_1, \theta_2)=|F_2^{\mathcal{O}}(\theta_1,\theta_2)|^2-|F_1^{\mathcal{O}}(\theta_1)|^2 |F_1^{\mathcal{O}}(\theta_2)|^2\,,
\label{9}
\eeq 
with $F_0^{\mathcal{O}}:=\bra 0|\mathcal{O}|0\ket$. 
For simplicity, I describe here a theory with a single particle spectrum, which means that the rapidities are sufficient to specify the state.  In IQFT these form factors satisfy consistency equations which can be solved exactly. This is known as the form factor bootstrap \cite{KW,smirnovbook}. Typically, the easiest form factors to compute are those associated to $k=1,2$, the one- and two-particle form factors. Often, the series above can indeed be truncated at two-particle order whilst providing valuable information. I will consider this particular truncation in this paper. 

There is one further step that is required, namely to extend these results to replica theories and BPTFs. Because of replication, a theory with a single particle spectrum, becomes an $n$-particle theory, which means that the form factors above should carry additional indices, indicating the copy number, and there will be additional sums over those indices in the form factor and cumulant expansions above. Because of the symmetry that $\TT$, $\tilde{\TT}$ implement, their form factor equations are different from those satisfied by more standard local fields. These equations were first written and solved for some theories in \cite{entropy}. I list some of the properties of one- and two-particle form factors below. For spinless fields such as the BPTFs, the form factors can only depend on rapidity differences (by Lorenz invariance). This means that the one-particle form factors are independent of rapidities and can be written as
\beq 
F_1^{\TT|j}(n)\quad \mathrm{with}\quad j=1,\ldots,n\,,
\eeq 
and the two-particle form factors are functions 
\beq 
F_2^{\TT|jk}(\theta;n)\quad \mathrm{with}\quad j,k=1,\ldots,n\,.
\eeq 
of a single rapidity $\theta$ and the replica number $n$ with labels $i,j$ indicating the copy number where the corresponding particle lives. 

Due to replica permutation symmetry and the monodromy properties of the form factors specified by the form factor equations \cite{entropy}, the $n$ one-particle form factors and the $n^2$ two-particle form factors above are not independent. In fact, since all copies are identical, all one-particle form factors  must be identical too 
\beq 
F_1^{\TT|j}(n)=F_1^{\TT|1}(n):=F_0^{\TT}(n) g_n\,,
\eeq
for all $j$, where $F_0^{\TT}(n):={}_n\bra 0|\TT|0\ket_n$ is the vacuum expectation value (VEV) in the replica theory and $g_n$ is therefore the ratio of one-particle form factor and VEV. Similarly, all two-particle form factors can be expressed in terms of a single independent function to which all others are related.  Since all copies are identical, it is enough to consider the functions $F_2^{\TT|1 j}(\theta;n)$ with $j>1$. They satisfy
\beq 
F_2^{\TT|1j}(\theta;n)=F_2^{\TT|1 1}(-\theta+ 2\pi i (j-1);n) \quad \mathrm{and} \quad F_2^{\tilde{\TT}|1 j}(\theta;n)=F_2^{\TT|1 1}(\theta+ 2\pi i (j-1);n)\,,
\eeq
It is then useful to define
\beq 
F_2^{\TT|1 1}(\theta;n):=F_0^{\TT}(n) f_n(\theta)\,.
\eeq 
The function $f_n(\theta)$ has the following monodromy properties,
\beq 
f_n(\theta)=S(\theta)f_n(-\theta)=f_n(-\theta+2\pi i n)\,, 
\label{MFF}
\eeq 
which are a consequence of the form factor equations for two-particle form factors. The function $S(\theta)$ is the two-particle scattering amplitude. A further important property concerns the pole structure of $f_n(\theta)$: it has kinematic poles at $\theta=i\pi$ and $\theta=i\pi (2n-1)$.

We can now generalize the cumulant expansion to BPTFs by simply including extra sums over the copies. Consider the following quantity
\beq 
C_n(t):=\log\left[\frac{{}_n\bra 0|\TT(0,0) \tilde{\TT}(0,t) |0\ket_n}{{}_n\bra 0|\TT|0\ket_n^2}\right]
\label{ratio}
\eeq 
where $|0\ket_n$ is the replica ground state. The first two contributions to the cumulant expansion of $C_n(t)$ are
\beq 
C_n^{(1)}(t)=\frac{n}{\pi}  |g_n|^2 K_0(i m t)\,.
\label{11}
\eeq 
\beq 
C_n^{(2)}(t)=\frac{n}{(2\pi)^2} \int_{-\infty}^\infty d\theta \left[\sum_{j=0}^{n-1} \left[f_{n}(\theta+2\pi i  j) f_{n}(-\theta+2\pi i j)- |g_n|^4\right]]\right] K_0(2 m i t \cosh \frac{\theta}{2})\,.
\label{12}
\eeq 
Both formulae are obtained by performing one integral, which for the one particle contribution immediately gives a modified Bessel function with imaginary argument, while for the two-particle contribution still leaves one integral. 

The temporal R\'enyi entropy (\ref{rentime}) is given by
\beq 
S_n(t)=\frac{C_n(t)}{1-n}+\frac{1}{1-n}\log(\varepsilon^{4\Delta_n}{}_n\bra 0|\TT|0\ket_n^2)\,,
\label{def}
\eeq 
and in the two-particle approximation we can just replace $C_n(t)$ by $C_n^{(1)}(t)+C_n^{(2)}(t)$. Already at this stage, if we compare to existing results for the spacial case, we see that the only difference is the replacement of the scale $m r$ by the scale $imt$, which amounts to analytic continuation to complex values of $m r$. Hence, apart from the constant term in (\ref{def}), the large size behaviour of the spacial R\'enyi entropy is characterized by exponential decay, whereas the leading large time behaviour of its temporal version is characterized by {\it damped oscillations}. I will discuss this in more detail below for the temporal von Neumann entropy of generic theories and for the temporal R\'enyi entropies of the free fermion theory. 
\section{Temporal von Neumann Entropy}
\label{von}
The temporal von Neumann entropy can be obtained as the limit $\lim_{n\rightarrow 1} S_n(t)$ of the R\'enyi entropies above. The index $n$ is the replica number, hence a positive integer greater than 1 and the cumulants typically involve sums like the one in (\ref{12}). This means that the limit $n\rightarrow 1$ is non-trivial. It is necessary to analytically continue the cumulants to $n \in \mathbb{R}^+$ first, a problem that was solved in \cite{entropy} for the second cumulant and more generally for all cumulants in free theories in \cite{nexttonext, davide,Castro-Alvaredo:2023wlr}. 
In \cite{entropy} the following identity (cotangent trick) was used:
\beq 
\sum_{j=0}^{n-1} s_n(\theta,j)=\frac{1}{2\pi i }\oint d z \pi \cot (\pi z) \, s_n(\theta,z)-\pi \sum_i \cot (\pi z_i)  \, \, \mathrm{Res}(s_n(\theta,z=z_i))\,, \label{sum}
\eeq 
where $s_n(\theta,z)=f_{n}(\theta+2\pi i z) f_{n}(-\theta+2\pi i z)$, $z_i$ are poles of $s_n(\theta,z)$ that fall within the integration contour. Taking the contour as a rectangle with corners at $\pm i L$ and $\pm iL+ n$, $L\rightarrow \infty$ ensures that the cotangent has exactly $n$ poles inside the contour whose residue contributions reproduce the original sum. Then using the property that the function $f_n(\theta)$ has kinematic poles at $\theta=i\pi$ and at $\theta=i \pi(2n-1)$, we find that the function $s_n(\theta,z)$ has four poles within the contour, namely
\beq 
z_1=\frac{1}{2}-\frac{\theta}{2\pi i}, \quad z_2=n-\frac{1}{2}-\frac{\theta}{2\pi i}, \quad
z_3=\frac{1}{2}+\frac{\theta}{2\pi i}, \quad z_4={n}-\frac{1}{2}+\frac{\theta}{2\pi i}.
\eeq 
Using the quasi-periodicity property of the form factors in (\ref{MFF}) and the periodicity of the cotangent one can show that
\beqa 
-\pi \sum_{i=1}^4 \cot (\pi z_i)  \, \, \mathrm{Res}(s_n(\theta,z=z_i)=-i 
\tanh\frac{\theta}{2} \left[f_n\left( i\pi + 2\theta \right) - f_n\left(  i \pi  - 2\theta\right)\right].
\eeqa 
What remains are the non-vanishing contributions from the contour integral. It is possible to argue that the contributions from the horizontal lines in the contour will vanish when $L\rightarrow \infty$ due to exponential decay of $s_n(\theta,z)$ and the vertical line contributions can be combined into a single integral due to the quasi-periodicity of $s_n(\theta,z+n)= S(\theta+2\pi i z)S(\theta-2\pi i z) s_n(\theta,z) $. This gives the full analytic continuation of the sum (\ref{sum}) as
\beqa 
{\mathcal{S}}(\theta,n) &=& -i 
\tanh\frac{\theta}{2} \left[f_n\left( i\pi + 2\theta \right) - f_n\left(  i \pi  - 2\theta\right)\right] \nonumber\\
&& -\frac{1}{4\pi i} \int_{-\infty}^{\infty} d\beta \,  \coth\frac{\beta}{2}(S(\theta+\beta)S(\theta-\beta)-1) s_n\left(\theta,\frac{\beta}{2\pi i}\right)\,. 
\label{firstline}
\eeqa 
This function has several interesting properties. The most important property is its limit $n\rightarrow 1$. In this limit all form factors are vanishing because the BPTF is trivial in the absence of replicas. This means that, for any non-zero values of $\theta$, the sum goes to zero when $n\rightarrow 1$. This also applies to the one-particle contributions to the cumulants (\ref{11}) and (\ref{12}). Non-trivial results are obtained however when we observe that the first line of (\ref{firstline}) behaves as a distribution around $\theta=0$. In other words, the simultaneous limits $n\rightarrow 1$ and $\theta\rightarrow 0$ generate a non-trivial contribution. In \cite{entropy} it was shown that
\beq 
\lim_{n\rightarrow 1} \frac{n}{1-n}{\mathcal{S}}(\theta,n) =-\frac{\pi^2}{2}\delta(\theta)\,,
\label{delta}
\eeq 
whereas one-particle form factors give no contribution in this limit. Although all the results of this section are a retelling of the known derivation \cite{entropy}, it is remarkable that the same arguments apply for the timelike cumulants. It also means that, by plugging  (\ref{delta}) into (\ref{12}) the $\theta$ integral can be computed exactly and we find that the temporal von Neumann entropy is given by 
\beq 
S(t)=-\frac{c}{3}\log{m \varepsilon}-U-\frac{1}{8} K_0(2 m i t)+  \cdots\,, 
\label{main}
\eeq 
where $U$ is a constant resulting from the limit $n\rightarrow 1$ of the vacuum expectation value contribution in (\ref{def}).
 Higher particle form factors will produce further oscillatory corrections with frequencies greater or equal than $3m$.
This result is identical to that obtained in \cite{entropy} under analytic continuation $r \mapsto it$ with $r,t \in \mathbb{R}$. This means that the temporal entropy receives oscillatory contributions which are damped as 
$t^{-\frac{1}{2}}$ for large times and which make the entropy complex. A discussion, generalization and interpretation of this result can be found in the conclusion.  Further analysis of the damping factor is given in Section~\ref{extra}.

\section{Oscillations and Damping}
\label{Free}
A feature of the analytic continuation (\ref{firstline}) is that the integral part is vanishing for free theories, that is, when the scattering matrix $S(\theta)=\pm 1$. This means that we can have a very explicit formula, up to the second cumulant for the R\'enyi entropies too. In this section I will consider this formula for the free fermion theory. In this case the function $f_n(\theta)=f_n^{\rm FF}(\theta)$ is given by \cite{entropy,nexttonext}
\beqa 
f^{\rm FF}_n(\theta)=\frac{-i \cos\frac{\pi}{2n} \sinh\frac{\theta}{2n}}{n\sinh\left(\frac{i\pi-\theta}{2n}\right) \sinh\left(\frac{i\pi+\theta}{2n}\right)}\,,
\eeqa
and the one particle form factor is zero for symmetry reasons. This means that the first cumulant $C_n^{(1) {\rm FF}}(t)=0$ and the second cumulant takes the form 
\beq 
C_n^{(2),{\rm FF}}(t)= -\frac{i  n}{(2\pi)^2} \int_{-\infty}^\infty d\theta  \,{\tanh\frac{\theta}{2}}\left(f^{\rm FF}_n(2\theta+i\pi)+f^{\rm FF}_n(2\theta-i\pi)\right) K_0(2 m i t \cosh \frac{\theta}{2})\,.
\label{122}
\eeq 
With these explicit formulae at hand, it is possible to obtain the leading large time asymptotics of $C_n^{(2),{\rm FF}}(t)$ by stationary phase analysis, similar to the analysis performed in \cite{viti2}.  Expanding around $\theta=0$ we get
\beq 
-i \,{\tanh\frac{\theta}{2}}\left(f^{\rm FF}_n(2\theta+i\pi)+f^{\rm FF}_n(2\theta-i\pi)\right)\approx \frac{1}{2}+\mathcal{O}(\theta^2)\,.
\eeq 
At the same time, for large times, we can expand the modified Bessel function:
\beq 
K_0(2imt \cosh\frac{\theta}{2})\approx \frac{\sqrt{\pi}e^{2mit  \cosh\frac{\theta}{2}}}{\sqrt{4 i mt \cosh\frac{\theta}{2}}}\,,
\eeq 
and for $\theta \sim 0$ we can further approximate to 
\beq 
K_0(2imt \cosh\frac{\theta}{2})\approx \frac{\sqrt{\pi}e^{2mit (1+ \frac{\theta^2}{8})}}{\sqrt{4 i mt}}\,.
\eeq 
We then find that for large times, the second cumulant contribution scales as
\beq 
C_n^{(2),{\rm FF}}(t)\sim \frac{n}{8\pi^{\frac{3}{2}}} \frac{e^{i (2mt+\frac{\pi}{2})}}{mt}\,.
\label{leadings}
\eeq 
The presence of an oscillatory part of frequency $2m$ and of damping by a power of $t$ are similar features as found for free fermions after a global mass quench \cite{viti2}, except that in the quench case the result is real (the exponential is replaced by a cosine) and the damping stronger ($t^{-\frac{3}{2}}$). 

It is interesting to track the origins of each contribution:

\begin{itemize}
\item The oscillatory part comes in both cases from the exponential $e^{-it e(\theta)}$ in the form factor expansion of the correlation function. It is independent of any details pertaining to the theory or the state, hence shared by the entanglement entropy after a quench and by the temporal entropy. 
\item The damping however is dependent on the choice of state and on the form factors, so it will vary if we change the fields in the correlator and/or the state. For instance, for the global mass quench the state is characterized by a special function $\hat{K}(\theta)$ \cite{Sotiriadis} and the stationary phase analysis around $\theta=0$ is affected by the properties of this function. This changes means that, also in our case, we would likely get a different power law if we did our calculation for a different state instead of the ground state we consider here. 
\item For any theory that has non-vanishing one-particle form factors, there will be a less strongly damped oscillatory contribution of lower frequency $m$, coming from the first cumulant (\ref{11}) and scaling as $t^{-\frac{1}{2}}e^{i m t}$. If the spectrum contains a single particle of mass $m$ or, in more general theories, if $m$ is the lightest particle in the spectrum, the oscillation frequency $m$ will only be seen in the R\'enyi entropies but never in the von Neumann entropy.

\item In summary, the results of the past two sections tell us that the lowest oscillation frequency we will be able to detect from the temporal von Neumann entropy is $2m$, while for the temporal R\'enyi entropies it will be either $m$ (if the one-particle form factor is non-zero) and $2m$ if it is zero. It is interesting to ask whether these lower frequency contributions also characterize the large-time asymptotics, which is governed by power-law decay. We address this question in detail in the next Section.

\end{itemize}

\medskip 

It is also interesting to observe that while in the quench computation, the leading oscillatory part comes from a form factor contribution $f_n(2\theta)$, here the functions that play a role are $f_n(2\theta\pm i\pi)$. For the global quench, the function $f_n(2\theta)$ arises from the structure of the coherent post-quench state, built on quasiparticle pairs of opposite momenta. In the temporal entropy case the functions are $f_n(2\theta\pm i\pi)$ and we have again a quasiparticle pair structure where now the pairs are made out of one in-coming (past) and one out-going particle (future) (via crossing) of opposite momenta.

\section{Some Generalities on the Leading Large-Time Asymptotics}
\label{extra}
The cases discussed in Sections~\ref{von} and \ref{Free} point to the fact that the von Neumann and R\'enyi entropies admit large-time expansions characterized by terms of the form  $t^{-p} e^{- i k m t}$, where the values of $p$ and $k$ vary depending of the order of the form factor expansion. Above we have found that the von Neumann entropy always contains a universal term with $k=2$ and $p=1/2$ while the R\'enyi entropy can contain a term with $p=1/2$ and $k=1$ but in some cases, like for free fermions, such term is absent. An important question is: are these the leading terms in the large-time expansion of the correlation function? In this section I give a partial answer to this question. In essence there are certain properties that can be shown in great generality under some natural assumptions, while others are much more theory dependent.  

The most general conclusions can be drawn for the R\'enyi entropies which are directly proportional to the logarithm of the two-point function of BPTFs. The simple argument I present here is widely known (see e.g. \cite{smirnovbook,mussardobook}).

Consider the cumulant expansion \eqref{7} for the BPTFs. Applying stationary phase approximation we find that the stationary point corresponds to $\theta_j=0$ for all $j=1,\ldots k$ so, for large $t$ we have
\beq 
\log\left[\frac{\bra 0|\TT(0,0)\tilde{\TT}(0,t)|0\ket}{\bra 0|\mathcal{O}|0\ket^2} \right] \approx \sum_{k=1}^\infty \frac{1}{k!}\left[\prod_{i=1}^k \int_{-\infty}^\infty \frac{d\theta_i}{2\pi}\right] \hat{H}_k^{\TT}(\theta_1,\ldots,\theta_k) e^{-i t m (1+\frac{\theta_j^2}{2})}\,,
\label{77}
\eeq 
where $\hat{H}_k^{\TT}(\theta_1,\ldots,\theta_k)$ is the leading approximation of the cumulants around $\theta_j=0$ and, for simplicity, we include in this definition the sum over copy numbers. If the cumulant scales as a non-vanishing constant near the stationary point, the corresponding form factor contribution to the sum above will scale as 
\beq 
\frac{1}{k!}\left[\prod_{i=1}^k \int_{-\infty}^\infty \frac{d\theta_i}{2\pi}\right] \hat{H}_k^{\TT}(\theta_1,\ldots,\theta_k) e^{-i t m (1+\frac{\theta_j^2}{2})}\sim  \frac{\hat{H}_k^{\TT}(0,\ldots,0)}{(2\pi)^{\frac{k}{2}}k!}  \frac{ e^{-i k(mt+\frac{\pi}{4})}}{(mt)^{\frac{k}{2}}}\,.
\label{compare}
\eeq
This means in particular that, if the one-particle cumulant is non-zero (as in minimal $E_8$ Toda, for instance \cite{e8}), the leading large-time scaling will be characterized by a term of the form $e^{-imt} t^{-\frac{1}{2}}$. If this term is absent, the two-particle form factor will determine the asymptotics. If the expansion of the cumulant around zero rapidities contains a constant term, we will find leading scaling $e^{-2mit} t^{-1}$, as above for the free fermion (\ref{leadings}). On the other hand, if the leading term in the cumulant expansion around zero rapidities contains any even power of the rapidities, the power-law decay will be more rapid, as $e^{-2mit} t^{-p}$ with $p>1$.  

Therefore, we can say that the leading large-time asymptotics of R\'enyi entropies will display one the two following behaviours: (1) It will scale as ${e^{-imt}}{(mt)^{-\frac{1}{2}}}$ in the absence of internal symmetries that constraint the one-particle form factor to vanish; (2) It will scale as $e^{-2mi t}{(mt)^{-p}}$ with $p\geq 1$ if the one-particle form factor is vanishing. 

\medskip 

The argument above is based on the natural assumption that the cumulants are regular functions around zero values of the rapidities. There is a subtlety for BPTFs, namely  while their cumulants are regular for $n$ integer, things change in the limit $n\rightarrow 1$, which is required to compute the von Neumann entropy. For this reason, conclusions for the temporal von Neumann entropy are different from the above.  An example were the cumulants behave as a distribution around $\theta=0$ is precisely for the two-particle contribution to the von Neumann entropy, as shown in (\ref{delta}). The one-particle contribution always vanishes in the $n\rightarrow 1$ limit, so in the form factor expansion, the lowest frequency oscillations always come from the two-particle form factor through the contribution (\ref{main}). This contribution scales as $e^{-2mit} t^{-\frac{1}{2}}$ for large times. Comparing to (\ref{compare}) we see that the power of $t$ is smaller (in absolute value) than expected from stationary phase analysis.

Is the scaling $e^{-2mit} t^{-\frac{1}{2}}$ always the leading contribution for large time in the temporal von Neumann entropy? The answer will likely be theory-specific, even though, again, stationary phase analysis would never produce a decay that is any slower than $t^{-\frac{1}{2}}$. The special properties of the BPTF form factors and their analytic continuations make it difficult to come to a definite conclusion without further analysis. It would be interesting to try and answer this question at least for free theories where all form factors are known. I leave this for a future project.

\section{Conclusion}
\label{conc}
The main achievement and original aim of this paper has been to show that many of the known properties of the temporal entanglement entropy \cite{Carignano:2023xbz,Bou-Comas:2024pxf,Carignano:2024jxb,Carignano:2025egi} are easy to establish in the context of 1+1D (integrable) quantum field theory. They are natural consequences of the symmetry between space and time variables in 1+1D, standard features of dynamical correlators and specific features of the dynamical correlators of BPTFs. These properties are: oscillatory and power-law dependence (decay) for large times and complex values.
\medskip 

Quantitatively, the main general results of this paper are the following:  a universal formula for the temporal von Neumann entropy in the two-particle form factor approximation (\ref{main}) and for the temporal R\'enyi entropies in the presence of non-vanishing one-particle form factors. In fact, both results can be easily generalized to theories with multi-particle spectra by just including a sum over particle types.  Neither result actually requires integrability \cite{next}, as they rely solely on universal properties of the one- and two-particle form factors.  This means that the following results are general for 1+1D massive QFT. First,
\beq 
S(t)=-\frac{c}{3}\log{m_1 \varepsilon}-U-\frac{1}{8}\sum_{a=1}^\ell K_0(2 m_a i t)+  \cdots\,, 
\label{bessel}
\eeq 
where $\ell$ is the number of particle types and the ellipsis indicate further oscillatory corrections of frequencies greater or equal $3m_1$. The two particle contribution universally decays as $t^{-\frac{1}{2}}$ for large times. More work is needed to establish if and when this is also the leading contribution. This is identical to equation (1.4) in \cite{entropy} under analytic continuation $r \mapsto i t$ with $r,t \in \mathbb{R}$. The technical reason for the universality of this result are the special properties of BPTFs and their correlators, especially in the limit $n\rightarrow 1$.

Second, whenever the one-particle form factors of the BPTF are non-vanishing, the temporal R\'enyi entropies scale as
\beq 
S_n(t)=\frac{n}{\pi(1-n)}\sum_{a=1}^\ell |g_n^a|^2 K_0(im_a t)+\cdots
\label{def2}
\eeq 
where $g_n^a$ are the normalized one-particle form factors, as in (\ref{11}) but now generalized to different particle types. For large times, the one-particle contribution decays as $t^{-\frac{1}{2}}$ which is the leading large-time behaviour. For theories where $g_n^a=0$ the leading large-time contribution comes from the two-particle form factor and, as for the von Neumann entropy, contains oscillatory terms of frequencies $2m_a$. It decays as $t^{-p}$ with $p\geq 1$, depending on the model.

\medskip

The oscillatory and power-law scaling of the entropies with respect to time, is reminiscent of post-quench dynamics.  My main conclusions are:
\begin{itemize}
\item Temporal entropies, are generally complex, with both the real and imaginary parts, damped oscillatory functions. Complex values are a feature of the temporal entropy, pseudoentropy and timelike entropies, as previously observed in the literature \cite{Doi:2023zaf,Xu:2024yvf,Doi:2022iyj}. In the context of QFT these are all natural features of dynamical correlators.
\item My results are just the analytic continuation of the spacial entropy to complex values of interval length $r:=it$. While in the spacial case, the Bessel functions in the cumulant expansion give rise to exponentially decaying corrections to entropy saturation, in the temporal picture they give rise to damped oscillations on top of a saturation value, representing the entanglement between pairs of quasiparticles produced at different time regions. This is in the same spirit as the quasiparticle picture of global quenches. In the spacial case, the connection between the Bessel functions (\ref{bessel}) and the entanglement of quasiparticle pairs following a quench, was proposed in \cite{next} and did not require integrability.

\item For theories with many quasiparticle species, the various Bessel function contributions will now give rise to damped oscillations of different frequencies. In other words, the particle spectrum of the theory could be read from the Fourier transform of temporal entropy. While in the spacial case, heavier particles are hard to detect, as their contributions are exponentially suppressed, in the temporal case they contribute as oscillatory functions of higher frequencies. At the same time, other oscillatory contributions will also arise from higher particle form factors, as for global quenches \cite{viti1,viti2,SG2}.
\end{itemize}
Despite its simplicity, the viewpoint presented here gives a complementary picture of temporal entanglement in gapped systems and highlights some of its universal properties.  

\medskip 
\noindent {\bf Acknowledgments:} I would like to thank Alvise Bastianello for his invitation to attend the recent workshop {\it Threefold Quantum: Theory, Experiment, Computation Out of Equilibrium}, which was held in Paris 11-13 March 2026. Among many interesting talks, the talk by Luca Tagliacozzo ``{\it Characterizing Ergodicities through Temporal Entanglement}" provided inspiration for this project. I thank the referees of the paper for their thoughtful comments.

\bibliography{Ref}
\end{document}